\documentclass{article}
\usepackage[utf8]{inputenc}
\usepackage[margin=1in]{geometry}
\usepackage{slashed,graphicx}
\usepackage{amsthm}
\usepackage{amsmath}
\usepackage{float}
\usepackage{amssymb}
\usepackage[caption=false]{subfig}

\title{Radially excited ($n = 3$) charm mesons in heavy quark effective theory}
\author{Ritu Garg, Kundan Kumar, A. Upadhyay
 \\\small{\it School of Physics and Material Science},\\\small{\it Thapar University,
Patiala, Punjab-147004}\\\small{E-mail: ritugarg039@gmail.com,
alka.iisc@gmail.com}}

\begin{document}

\maketitle
\section{Abstract} By exploring heavy quark effective theory (HQET), we use theoretical available data for bottom mesons to analysis the masses and decays for n = 3 charm mesons. From the predicted masses, we studied ground state strong decay modes in terms of couplings. Comparing the decays with available total decay widths, we provide upper bounds on the associated couplings. We also plot Regge trajectories for our predicted data in planes (J, $M^2$ ) and ($n_r$, $M^2$ ) and estimated higher masses (n = 4) by fixing Regge slopes and intercepts. These Regge trajectories are used to clarify $D_2^*(3000)$ state's $J^P$ as 1F ($2^+$) state. The presented results may further get confirmation through upcoming experimental information.

\section{Introduction}
In recent decades, a noteworthy experimental development is achieved which explored spectrum of heavy light hadrons. Several new candidates like $D^*_{2}(3000)$, $D^{*}_{J}(3000)$,  $D_{J}(3000)$, $D^{*}_{3}(2760)$, $D^{*}_{1}(2680)$, $D^{*}_{2}(2460)$, $D^{*}_{J}(2760)$, $D_{S0}(2590)^+$ were observed by experiment facilities like LHCb, BABAR, BESIII etc., have flourished charm meson spectrum [1-5]. In case of bottom sector, many new states like $B_{J}(5840)$, $B_{J}(5960)$, $B_{S1}(5830)$, $B^{*}_{S2}(5840)$, $B_{S}^{0}(6063.5)$, $B_{S}^{0}(6108.8)$, $B_{S}^{0}(6114)$, $B_{S}^{0}(6158)$ have broaden the bottom spectrum [6-8]. There is also remarkable growth in the field of baryon sector with the observations of five narrow $\Omega_{c}$ resonances [9] and doubly charmed $\Xi_{cc}$ resonances [10]. However in this paper, we are focusing on mesons only.

The masses and decay widths of charm mesons for ground state (1S) and first excited state (1P) are well established experimentally and mentioned in PDG for both strange and non strange sector [11]. Recently LHCb collaboration studied $B^{-} \longrightarrow D^{+}\pi^{-}\pi^{+}$ decays with Dalitz plot analysis technique and reported the existence of charm resonances with spin 1, 2, 3 at $D^{+}\pi^{-}$ mass spectrum [1]. Their investigation found that these charm resoances are mainly coming from contribution of $D^{*}_{3}(2760)$, $D^{*}_{1}(2680)$, $D^{*}_{2}(2460)$ and $D^*_{2}(3000)$ charmed meson decays into S-wave $D^{+}\pi^{-}$. The obtained masses and decay widths for these resonances are: \\\\
$M(D_{2}^{*}(2460))= 2463.7\pm0.4(stat)\pm0.4(syst)$ MeV\\
$\Gamma(D_{2}^{*}(2460)) = 47\pm0.8\pm0.9$ MeV/$c^2$\\\\
$M(D_{1}^{*}(2680))= 2681.1\pm5.6(stat)\pm4.9(syst)$ MeV\\
$\Gamma(D_{1}^{*}(2680)) = 186.7\pm8.5\pm8.6$ MeV/$c^2$\\\\
$M(D_{3}^{*}(2760))= 2775.5\pm4.5(stat)\pm4.5(syst)$ MeV\\
$\Gamma(D_{3}^{*}(2760)) = 95.3\pm9.6\pm7.9$ MeV/$c^2$\\\\
$M(D_{2}^{*}(3000))= 3214\pm29(stat)\pm33(syst)$ MeV\\
$\Gamma(D_{2}^{*}(2460)) = 186\pm38\pm34$ MeV/$c^2$\\\\
In 2015, LHCb group examined $B^{0} \longrightarrow \Bar{D^{0}}\pi^{-}\pi^{+}$ decays and measured $D_{0}^{*}(2300)$ and $D_{0}^{*}(2460)$ charm meson state. They assigned state $D_{3}^{*}(2760)$ with $J^{P} = 3^{-}$ for the first time [2]. In year of 2013 and 2010, a remarkable achievement received by BABAR and LHCb groups [3-4]. LHCb (2013) detector discovered two resonances $D_{J}^{*}(2650)^{0}$, $D_{0}^{*}(2760)$ with natural parity and two resonances $D_{J}(2580)^{0}$ and $D_{J}(2740)^{0}$ with unnatural parity by analysing invariant mass spectrum of $D^{+}\pi^{-}$,  $D^{0}\pi^{+}$, $D^{*+}\pi^{-}$. In addition, state $D_{J}(3000)^{0}$ with unnatural parity and state $D_{J}^{*}(3000)^{0}$ with natural parity were found in $D^{*+}\pi^{-}$, $D^{+}\pi^{-}$ mass spectrum respectively. In 2010, BABAR experiment analyzed $e^{+}e^{-}$ inclusive collision at centre of mass energy 10.58 GeV and observed $D_{J}(2560)^{0}$, $D_{J}(2600)^{0}$, $D_{J}(2600)^{+}$, $D_{J}(2750)^{0}$, $D_{J}^{*}(2760)^{+}$ and $D_{J}(2760)^{0}$. Masses and decay widths of states confirmed by LHCb and BABAR are listed in Table I.\\

\begin{table}[ht]
    \centering
    \begin{tabular}{|c|c|c|c|c|c|c|c|}
    \hline
    States &  BABAR(2010)[4] & LHCb(2013)[3] & LHCb(2015)[2] & LHCb(2016)[1]  \\
    \hline
    $D_{1}(2420)^{0}$ & M: $2420.1 \pm{0.1}\pm{0.8}$& M: $2419.6\pm{0.1}\pm{0.8}$ & - &-\\
        & $ \Gamma =31.4\pm{0.5}\pm{1.3}$   &  $ \Gamma =35.2\pm{0.4}\pm{0.9}$  &-     &-\\
        \hline
         $D_{2}^{*}(2460)^{0}$ & M: $2462.2 \pm{0.1}\pm{0.8}$& M: $2460.4\pm{0.1}\pm{0.1}$ & - &-\\
         & $ \Gamma =50.5\pm{0.6}\pm{0.7}$   &  $ \Gamma =45.6\pm{0.4}\pm{0.1}$  & -    &-\\
        \hline
         $D(2550)^{0}$ & M: $2539.4 \pm{4.5}\pm{6.8}$& M: $2579.5\pm{3.4}\pm{5.5}$ &-  &-\\
        & $ \Gamma =130\pm{12}\pm{13}$   &  $ \Gamma =177.5\pm{17.8}\pm{46.0}$  &-     &-\\
        \hline
        $D_{J}^{*}(2600)^{0}$ & M: $2608.7 \pm{2.4}\pm{2.5}$&  M: $2649.2 \pm{3.5}\pm{3.5}$  &  & M: $2681.1 \pm{5.6}\pm{4.9}$\\
        & $ \Gamma =93\pm{6}\pm{13}$ &$ \Gamma =140.2\pm{17.1}\pm{18.6}$  & -    & $ \Gamma =186.7\pm{8.5}\pm{8.6}$\\
        \hline
         $D_{3}^{*}(2750)^{0}$ & M: $2763.3 \pm{2.3}\pm{2.3}$& M: $2760.1\pm{1.1}\pm{3.7}$ & - &M: $2775.5\pm{4.5}\pm{4.5}$ \\
        & $ \Gamma =60.9\pm{5.1}\pm{3.6}$   &  $ \Gamma =74.4\pm{3.4}\pm{19.1}$  &     & $ \Gamma =95.3\pm{9.6}\pm{7.9}$ \\
        \hline
        $D_{3}^{*}(2750)^{+}$ & M: $2769.7 \pm{3.8}\pm{1.5}$& M: $2771.7\pm{1.7}\pm{3.8}$ & - & -\\
        & $ \Gamma =60.9$   &  $ \Gamma =66.7\pm{6.6}\pm{10.5}$  &  -   &  -\\
        \hline
        $D_{2}^{*}(2460)^{+}$ & & M: $2463.1\pm{0.2}\pm{0.6}$& -   & -\\
        & -  &  $ \Gamma =48.6\pm{1.3}\pm{1.9}$  & -    & - \\
        \hline
         $D(2740)^{0}$ & & M: $2737.0\pm{3.5}\pm{11.2}$&-    & -\\
        &  - &  $ \Gamma =73.2\pm{13.4}\pm{25.0}$  &   -  & - \\
        \hline
         $D^{*}(3000)^{0}$ & & M: $2971.8\pm{8.7}$& -   &- \\
        &  - &  $ \Gamma =188.1\pm{44.8}$  &    - &  -\\
        \hline
        $D^{*}_{J}(3000)^{0}$ & & M: $3008.1\pm{4.0}$&   - &- \\
        & -  &  $ \Gamma =110.5\pm{11.5}$  &  -   &-  \\
        \hline
 $D^{*}_{2}(2460)^{-}$ & & & M: $2468.6\pm{0.6}\pm{0.3}$    & -\\
          & -  & -& $ \Gamma =47.3\pm{1.5}\pm{0.7}$      & - \\
        \hline
         $D^{*}_{3}(2750)^{-}$ & & & M: $2798\pm{7}\pm{1}$    & -\\
        & -  & - & $ \Gamma =105\pm{18}\pm{6}$       &  -\\
        \hline
         $D^{*}_{2}(3000)^{0}$ & & & & M: $3214\pm{29}\pm{33}$     \\
        & -  &  -& -& $ \Gamma =186\pm{38}\pm{34}$         \\
        \hline

         \end{tabular}
    \caption{States observed by different experimental facilities with masses in MeV and decay widths in $ MeV/C^2$}
    \label{tab:my_label}
\end{table}

Different  theoretical models like $^3P_{0}$ model [12], HQET [13], QCD sum rule [14], relativized quark model [15-16] examined all above mentioned states, computed their masses and suggested their $J^{P}$ values. States $D^{*}_{0}(2300)$, $D_{1}(2420)$, $D_{1}(2430)$ and $D^{*}_{2}(2460)$ are reported in PDG [11] and their assigned $J^P$ are $1^3P_{0}$, $1^1P_{1}$, $1^3P_{1}$, $1^3P_{2}$ respectively. Results of $D(2550)^{0}$ state observed by BABAR [4] are similar to those of the $D_{J}(2580)^{0}$ state reported by LHCb [1] and considered as candidate of $2^{1}S_{0}$ state. Information provided by LHCb group for the states $D_{1}^{*}(2680)$ in 2016 [1] and $D_{1}^{*}(2650)$ in 2013 [3] are similar to those of state $D^{*}(2600)$ observed by BABAR collaboration [4]. States $D_{1}^{*}(2680)$, $D^{*}(2600)$, $D_{1}^{*}(2650)$ are probably same particle and theoretical studies suggested as $2^{3}S_{1}$ state [17-25]. The mass and decay width of state $D_{3}^{*}(2750)$ reported by LHCb group in 2015 [2] and $D^{*}_{J}(2760)^{0}$ in 2013 [3] are close with results of $D^{*}(2760)^{0}$ observed by BABAR collaboration [4]. So all these states $D_{3}^{*}(2750)$, $D^{*}(2760)^{0}$, $D^{*}_{J}(2760)^{0}$ may be similar state and theoretical studies suggested them to be the candidate of 1D $3^{-}$ state [17-27]. Observations of the state $D_{J}(3000)$ with unnatural parity, $D^{*}_{J}(3000)$ with natural parity by LHCb [3] suggested  possible assignment for state $D_{J}^{*}(3000)$ are $3^3S_{1}$, $2^3P_{2}$, $1^3F_2$ and $1^3F_{4}$ and for state $D_{J}(3000)$ are $3^3S_{0}$, $2^3P_{1}$ respectively. LHCb detector observed the state $D_{2}^{*}(3000)^{0}$ with $J^{P}$ = $2^{+}$ [1] and suggested it to be a candidate of $3^3P_{2}$, $1^3F_2$ state. Theoretical approaches also examined $D_{2}^{*}(3000)^{0}$, $D_{J}^{*}(3000)$, $D_{J}(3000)$ states and tried to assign their proper $J^P$ state. In Ref [28] Zhi-Gang Wang et.al analyzed the state $D_{2}^{*}(3000)^{0}$ and assigned it 1F $2^{+}$. They also examined state $D_{2}^{*}(3000)^{0}$ with QPC (quark pair creation) model by studying decays of $3^3P_{2}$ and $2^3F_{2}$ charmed mesons. They suggested it to be  $3^3P_{2}$ state but possibility of $2^3F_{2}$ state was also not excluded. The states $D_{J}^{*}(3000)$ and $D_{J}(3000)$ also studied by S.Godfrey and K.Moats, identified them as $3^3P_{2}$, $1^3F_2$ respectively [29]. Authors in Ref [30] studied the state $D_{J}^{*}(3000)$, $D_{J}(3000)$ and suggested most favourable assignment for them to be 2P($0^+$, $1^+$) respectively. \\
With recent experimental information, we are motivated to look for other excited states. In this paper, we are predicting masses for n = 3 charm mesons in framework of heavy quark effective theory(HQET). We analyze possible ground state decay modes of predicted charm mesons. We examine the state $D_{2}^{*}(3000)^{0}$ using Regge trajectory and specify its spin parity state. Also, we study decays from excited state to ground state via a pseudoscalar mesons only and evaluate unknown coupling constants. We construct Regge trajectories for predicted masses in plane ($M^2$, J) and ($M^2$, $n_r$) and masses are estimated for higher charm states by fixing Regge slope and intercepts. This paper is summarized as follows: Section II gives brief description of HQET formalism and introduction of heavy quark symmetry parameters. Section III presents numerical analysis where we predicted masses for n = 3 charm mesons and their decays. Section IV gives conclusion of the paper.
\section{Theoretical Setup}
The study of excited charmed meson can be explored in framework of heavy quark effective theory (HQET). HQET is effective tool to describe properties of heavy light mesons like masses, decay widths, branching ratios, fractions, spin, parity etc [31]. This theory flourished with two approximate symmetries- heavy quark symmetry and chiral symmetry. Heavy quark symmetry is applicable in approximation $m_Q\to\infty$. In limit $m_Q\rightarrow \infty$, spin of light quarks decoupled from spin of heavy quark, so total angular momentum of light quarks remains conserved. Total  angular momentum of light quarks is $s_{l}$ = $s_{q}$+$l$, $s_{q}$ = spin of light quark(1/2) and $l$ = total orbital momentum of light quarks. In heavy quark limit, mesons are classified in doublets on basis of total angular momentum $s_{l}$ of light quarks. For $l = 0$, $s_{l}$ = 1/2 combining this with spin of heavy quark $s_{Q}$ = 1/2 and resulted with doublet $(0^{-},1^{-})$. This doublets is represented by $(P,P^{*})$. $l = 1$ corresponds to  two doublets represented by $(P^{*}_{0}, P_{1}^{'})$ and $(P_{1}, P_{2}^{*})$ with $J_{s_{l}}^{P} = (0^{+}, 1^{+})$ and $J_{s_{l}}^{P} = (1^{+}, 2^{+})$ respectively. For $l = 2$, there are two doublets denoted by $(P^{*}_{1}, P_{2})$ and $(P_{2}^{'},P_{3}^{*})$ having $J_{s_{l}}^{P} = (1^{-}, 2^{-})$ and $J_{s_{l}}^{P} = (2^{-}, 3^{-})$ respectively. These doublets are expressed by super effective fields $H_{a}, S_{a}, T_{a}$ [19, 32-33] and expression for fields are given below:
\begin{gather}
\label{eq:lagrangian}
 H_{a}=\frac{1+\slashed
v}{2}\{P^{*}_{a\mu}\gamma^{\mu}-P_{a}\gamma_{5}\}\\
S_{a} =\frac{1+\slashed v}{2}[{P^{'\mu}_{1a}\gamma_{\mu}\gamma_{5}}-{P_{0a}^{*}}]\\
T^{\mu}_{a}=\frac{1+\slashed v}{2}
\{P^{*\mu\nu}_{2a}\gamma_{\nu}-P_{1a\nu}\sqrt{\frac{3}{2}}\gamma_{5}
[g^{\mu\nu}-\frac{\gamma^{\nu}(\gamma^{\mu}-\upsilon^{\mu})}{3}]\}
\end{gather}

Here $H_{a}$ field belongs to $s_{l}^{p} = \frac{1}{2}^{-}$; $S_{a}$ and $ T_{a}$ have $s_{l}^{p} = \frac{1}{2}^{+}$ and $s_{l}^{p} = \frac{3}{2}^{+}$ respectively. For radial quantum number n = 2, these states are represented by a tilde over notations as showing $\Tilde{P}$, $\Tilde{P^{*}}$ and so on. For n = 3, these states are denoted by $\Tilde{\Tilde{P}}$, $\Tilde{\Tilde{P^{*}}}$ and so on. Here \textit{a} is light quark($u, d, s$) flavor index. $v^{\mu}$ is heavy quark four velocity and is conserved in strong interactions. The approximate chiral symmetry $SU(3)_L\times SU(3)_R$ is incorporated by field of pseudoscalar mesons ($\pi, \eta, k $). These pseudoscalar mesons are considered as approximate goldstone bosons and described by matrix field $\xi=e^{\frac{i\mathcal{M}}{f_{\pi}}}$ and $\Sigma=\xi^{2}$. Here $f_{\pi}$ is pion constant of 130 MeV and $\mathcal{M}$ is expressed as:
\begin{center}
\begin{equation}
\mathcal{M} = \begin{pmatrix}
\frac{1}{\sqrt{2}}\pi^{0}+\frac{1}{\sqrt{6}}\eta & \pi^{+} & K^{+}\\
\pi^{-} & -\frac{1}{\sqrt{2}}\pi^{0}+\frac{1}{\sqrt{6}}\eta &
K^{0}\\
K^{-} & \overline{K}^{0} & -\sqrt{\frac{2}{3}}\eta
\end{pmatrix}
\end{equation}
\end{center}
By including super fields $H_{a}, S_{a}, T_{a}$ defined in equations (1-3) and fields of goldstone bosons $\Sigma$, effective lagrangian of heavy light mesons is given as [19]:
\begin{multline}
    \mathcal{L} = iTr[\bar{H}_{b}v^{\mu}D_{\mu ba}H_{a}] +  \frac{f_\pi^{2}}{8}Tr[\partial^{\mu}\Sigma\partial_{\mu}\Sigma^{+}] + Tr[\bar{S_{b}}(iv^{\mu}D_{\mu ba} - \delta_{ba} \Delta_{S})S_{a}] + Tr[\bar{T_{b}^{\alpha}}(iv^{\mu}D_{\mu ba}\\- \delta_{ba}\Delta_{T})T_{a \alpha}] + \nonumber 
            Tr[\bar{X_{b}^{\alpha}}(iv^{\mu}D_{\mu ba} - \delta_{ba}\Delta_{X})X_{a \alpha}]   
\end{multline}

Here operator $\textit{D}_\mu$ is chirally covariant and expressed as $D_{\mu ab}= -\delta_{ab}\partial_{\mu}+\mathcal{V}_{\mu ab}$ 
= $-\delta_{ab}\partial_{\mu}+\frac{1}{2}(\xi^{+}\partial_{\mu}\xi+\xi\partial_{\mu}\xi^{+})_{ab}$. Here $\Delta_{F}$ is mass parameter give the mass difference between higher mass doublets (F) and lowest lying doublet (H) in terms of spin average masses of these doublets with same principle quantum number (n). This mass parameter can be described in terms of spin average mass of these doublets as [19]:
\begin{align}
             \Delta_{F}=\overline{M_{F}}&- \overline{M_{H}},~~ F= S,T,X,Y\\
\text{where, }~~~~~~~~~~~
           \overline{M_{H}}&=(3m^{Q}_{H^*}+m^{Q}_{H})/4\\
         \overline{M_{S}}&=(3m^{Q}_{S^*}+m^{Q}_{H})/4\\
         \overline{M_{T}}&=(5m^{Q}_{T^*}+3m^{Q}_{T})/8 
         \end{align}
          In heavy quark limit, mass degeneracy between members of meson doublets breaks and specific lagrangian for mass terms are:
           \begin{multline}
          \mathcal{L}_{1/m_{Q}} = \frac{1}{2m_{Q}}[\lambda_{H} Tr(\overline H_{a}\sigma^{\mu\nu}{H_{a}}\sigma_{\mu\nu}) + \lambda_{S}Tr(\overline S_{a}\sigma^{\mu\nu} S_{a}\sigma_{\mu\nu})+\lambda_{T}Tr(\overline T_{a}^{\alpha}\sigma^{\mu\nu}{T_{a}^{\alpha}}\sigma_{\mu\nu})]
          \end{multline} 
          Here parameters $\lambda_{H}$, $\lambda_{S}$, $\lambda_{T}$ are analogous with hyperfine splittings and defined as in Eq.(10-12). This mass terms in lagrangian represent only first order in $1/m_{Q}$ terms, but higher order terms may also be present otherwise. We are limiting to the first order corrections in $1/m_{Q}$.
 \begin{gather}
     \lambda_{H} = \frac{1}{8}(M^{2}_{P^{*}} - M^{2}_{P}) \\
     \lambda_{S} = \frac{1}{8}({M^{2}_{P^{'}}} - {M^{2}_{P^*}})\\
     \lambda_{T}=\frac{3}{16}({M^{2}_{P_2^*}}-{M^{2}_{P_1}})
     \label{100}
 \end{gather}
  Here we are motivated by fact that at scale of 1 GeV, when we study HQET, flavour symmetry spontaneously arises for b (bottom quark) and c (charm quark) and hence elegance of flavor symmetry refers to
        \begin{align}
           \label{1eu_eqn}
           \Delta_{F}^{(c)} =\Delta_{F}^{(b)}\\
   \lambda_{F}^{(c)} = \lambda_{F}^{(b)}
   \label{2eu_eqn}
\end{align}
Two body strong interaction through light pseudoscalar meson can be derived from heavy meson chiral lagrangian $L_{HH}$, $L_{SH}$, $L_{TH}$ and these interaction terms are written as [34-38]:
\begin{center}
\begin{gather}
\label{eq1:lagrangian}
L_{HH}={g_{HH}}Tr\{\overline{H}_{a}
H_{b}\gamma_{\mu}\gamma_{5}A^{\mu}_{ba}\}\\ L_{SH}=g_{SH}Tr\{\overline{H}_{a}S_{b}\gamma_{\mu}\gamma_{5}A^{\mu}_{ba}\}+h.c.\\
L_{TH}=\frac{g_{TH}}{\Lambda}Tr\{\overline{H}_{a}T^{\mu}_{b}(iD_{\mu}\slashed
A + i\slashed D A_{\mu})_{ba}\gamma_{5}\}+h.c.
\end{gather}
\end{center}

From these lagrangians, we can determine strong decay widths expressions for heavy light meson decays to ground state along with light pseudoscalar mesons M ($\pi,\eta,K$). These expression are described as:\\\\
   $(0^{-},1^{-}) \rightarrow (0^{-},1^{-}) + M$($\pi,\eta,K$)
\begin{gather}
\label{eq2:lagrangian}
 \Gamma(1^{-} \rightarrow 1^{-})=
C_{M}\frac{g_{HH}^{2}M_{f}p_{M}^{3}}{3\pi f_{\pi}^{2}M_{i}}\\
\Gamma(1^{-} \rightarrow 0^{-})=
C_{M}\frac{g_{HH}^{2}M_{f}p_{M}^{3}}{6\pi f_{\pi}^{2}M_{i}}\\
\Gamma(0^{-} \rightarrow 1^{-})=
C_{M}\frac{g_{HH}^{2}M_{f}p_{M}^{3}}{2\pi f_{\pi}^{2}M_{i}}
\end{gather}\\
 $(0^{+},1^{+}) \rightarrow (0^{-},1^{-}) + M$
\begin{gather}
\label{eq3:lagrangian}
 \Gamma(1^{+} \rightarrow 1^{-})=
C_{M}\frac{g_{SH}^{2}M_{f}(p^{2}_{M}+m^{2}_{M})p_{M}}{2\pi f_{\pi}^{2}M_{i}}\\
\Gamma(0^{+} \rightarrow 0^{-})=
C_{M}\frac{g_{SH}^{2}M_{f}(p^{2}_{M}+m^{2}_{M})p_{M}}{2\pi
f_{\pi}^{2}M_{i}}
\end{gather}\\
 $(1^{+},2^{+}) \rightarrow (0^{-},1^{-}) + M$
\begin{gather}
\label{eq4:lagrangian}
 \Gamma(2^{+} \rightarrow 1^{-})=
C_{M}\frac{2g_{TH}^{2}M_{f}p_{M}^{5}}{5\pi f_{\pi}^{2}\Lambda^{2}M_{i}}\\
\Gamma(2^{+} \rightarrow 0^{-})=
C_{M}\frac{4g_{TH}^{2}M_{f}p_{M}^{5}}{15\pi f_{\pi}^{2}\Lambda^{2}M_{i}}\\
\Gamma(1^{+} \rightarrow 1^{-})=
C_{M}\frac{2g_{TH}^{2}M_{f}p_{M}^{5}}{3\pi
f_{\pi}^{2}\Lambda^{2}M_{i}}
\end{gather}\\
where $M_{i}$, $M_{f}$ represents initial and final momentum, $\Lambda$ is chiral symmetry breaking scale of 1 GeV. $p_{M}$, $m_{M}$ denotes to final momentum and mass of light pseudoscalar meson. Coupling constant plays key role in phenomenology study of heavy light mesons. These dimensionless coupling constants describes strength of transition between H-H field (negative-negative parity), S-H field (positive-negative parity), T-H field (positive-negative parity). For transition from n = 3 to n = 1 coupling constants are given by $\tilde{\tilde{g}}_{HH}$, $\tilde{\tilde{g}}_{SH}$, $\tilde{\tilde{g}}_{TH}$  etc. and transition from n = 3 to n = 2 is denoted by $\tilde g_{HH}$, $\tilde{g}_{SH}$, $\tilde{g}_{TH}$. The coefficient $C_{M}$ for different pseudoscalar particles are:
$C_{\pi^{\pm}}$, $C_{K^{\pm}}$, $C_{K^{0}}$, $C_{\overline{K}^{0}}=1$, $C_{\pi^{0}}=\frac{1}{2}$ and $C_{\eta}=\frac{2}{3}(c\bar{u}, c\bar{d})$ or $\frac{1}{6}(c\bar{s})$. In our paper, we are not including higher order corrections of $\frac{1}{m_{Q}}$ to bring new couplings. We also expect that higher corrections give negligible contribution in comparison of leading order contributions.
\section{Numerical Analysis:}
Several higher charm states like  $D^*_{2}(3000)$, $D^{*}_{J}(3000)$, $D_{J}(3000)$, $D^{*}_{3}(2760)$, $D^{*}_{1}(2680)$, $D^{*}_{2}(2460)$, $D^{*}_{J}(2760)$ were discovered by LHCb and BABAR and analysed with different theoretical models. Theoretical studies with different theoretical approaches give different assignment to these states. Therefore, it is important to have better theoretical understanding for higher charm states. So in this paper, we aim to compute masses and decay width for n = 3 S-wave and P-wave charm mesons with their strange partners. Masses and decay widths for n = 2 charm spectra is already calculated with same framework.

 \subsection{Masses}
Mass is most important parameter to understand spectroscopy of heavy light mesons. To calculate masses for n = 3 S-wave and P-wave charm mesons, firstly we determine average masses $\overline{M_{F}}$, then compute symmetry parameters  $\Delta_{F}$, $\lambda_{F}$  for input values listed in Table II.
\begin{table}[ht]
        \centering
       \caption{Input values used in this work [39]. All values are in units of MeV.}
          \begin{tabular}{|c|c|c|c|c|c|}
      \hline
     State & $J^{P}$ & $ b\overline{q}$ & $ b\overline{s}$ & $ c\overline{q}$ & $ c\overline{s}$ \\
     \hline
     $3^{1}S_{0}$ & $0^{-}$ & 6362.22 & 6463.07 & - & -  \\
     \hline
      $3^{3}S_{1}$ & $1^{-}$ & 6342.74 & 6474.55 & - & - \\
      \hline
      $3^{3}P_{0}$ & $0^{+}$ & 6629 & 6731 & - & -  \\
      \hline
      $3^{1}P_{1}$ & $1^{+}$ & 6685 & 6768 & - & -  \\
      \hline
      $3^{3}P_{1}$ & $1^{+}$ &6650  & 6761 & - & -  \\
      \hline
      $3^{3}P_{2}$ & $2^{+}$ &6678  & 6780 & - & -  \\
      \hline
       $2^{1}S_{0}$ & $0^{-}$ & 5890 & 5976 & - & -  \\
     \hline
      $2^{3}S_{1}$ & $1^{-}$ & 5906 & 5992 & - & - \\
      \hline
       $2^{3}P_{0}$ & $0^{+}$ & 6221 & 6318 & - & -  \\
      \hline
      $2^{1}P_{1}$ & $1^{+}$ & 6281 & 6345 & - & -  \\
      \hline
      $2^{3}P_{1}$ & $1^{+}$ &6209  & 6321 & - & -  \\
      \hline
      $2^{3}P_{2}$ & $2^{+}$ &6260  & 6359 & - & -  \\
      \hline
      $2^{1}S_{0}$ & $0^{-}$ &  &  & 2581 & 2688  \\
     \hline
      $2^{3}S_{1}$ & $1^{-}$ & - & - & 2632 & 2731 \\
      \hline
       $2^{3}P_{0}$ & $0^{+}$ &- & - & 2919 & 3054  \\
      \hline
      $2^{1}P_{1}$ & $1^{+}$ & - & - & 3021 & 3154  \\
      \hline
      $2^{3}P_{1}$ & $1^{+}$ & - & - &  2932 & 3067  \\
      \hline
      $2^{3}P_{2}$ & $2^{+}$ &-  & - & 3012 & 3142   \\
      \hline
     \end{tabular}
    \label{tab:my_label1}
\end{table}
 The symmetry parameters for excited states can be expressed as:
\begin{gather}
\Delta_{\tilde{\tilde{H}}} = \overline M_{\tilde{\tilde{H}}}- \overline M_{\tilde{H}}\\
\Delta_{\tilde{\tilde{S}}} = \overline M_{\tilde{\tilde{S}}}- \overline M_{\tilde{\tilde{H}}}\\ 
\Delta_{\tilde{\tilde{T}}} = \overline M_{\tilde{\tilde{T}}}- \overline M_{\tilde{\tilde{H}}}\\ 
 \lambda_{\tilde{\tilde{H}}} = \frac{1}{8}(M^{2}_{\tilde{\tilde{P^{*}}}} - M^{2}_{\tilde{\tilde{P}}}) \\
     \lambda_{\tilde{\tilde{S}}}  =  \frac{1}{8}(M^{2}_{\tilde{\tilde{P^{'}}}} - M^{2}_{\tilde{\tilde{P^{*}}}} )\\
     \lambda_{\tilde{\tilde{T}}} =\frac{3}{16}({M^{2}_{\tilde{\tilde{P_2^*}}}}-{M^{2}_{\tilde{\tilde{P_1}}}})
\end{gather}
So, by help of heavy quark symmetry $\Delta_{F}^{(c)} =\Delta_{F}^{(b)}$, $\lambda_{F}^{(c)} = \lambda_{F}^{(b)}$ and using calculated flavor symmetry parameters $\Delta_{\tilde{\tilde{H}}}$, $\Delta_{\tilde{\tilde{S}}}$,  $\lambda_{\tilde{\tilde{H}}}$, $\lambda_{\tilde{\tilde{S}}}$, $\lambda_{\tilde{\tilde{T}}}$ we obtain masses for $n = 3$ S-wave and P-wave charm spectra. Obtained results are listed in Table III.
\begin{table}[ht]
\centering
\caption{Predicted masses for radially excited charm mesons}
 \begin{tabular}{| c | c | c | c | c | c | c |}
      \hline
      \multicolumn{1}{|c|}{} & \multicolumn{6}{c|}{Masses of n = 3 charm Mesons (MeV)}\\
       \cline{2-7}
     
       \multicolumn{1}{|c|}{$J^{P}(n^{2S+1}L_{J})$}&\multicolumn{3}{c|}{Non-Strange}&\multicolumn{3}{c|}{Strange}\\
       \cline{2-7}
       & \multicolumn{1}{c|}{Predicted}&\multicolumn{1}{c|}{[39]}&\multicolumn{1}{c|}{[40]}&\multicolumn{1}{c|}{Predicted}&\multicolumn{1}{c|}{[39]}&\multicolumn{1}{c|}{[40]}\\
       \hline
        $0^{-}(3^1S_0)$ &  3030.09 &  3062 & 2904& 3186.5 & 3219 & 3044\\
        \hline
        $1^{-}(3^3S_1)$ &  3064.45 &  3096 &  2947 & 3209.74 & 3242 & 3087 \\
        \hline
        $0^{+}(3^3P_0)$ &  3243.17 & 3346  & 3050& 3496.46 & 3541 & 3214\\
        \hline
        $1^{+}(3^1P_1)$ &  3356.13 &  3461 &  3082& 3567.18 & 3618 & 3234\\
        \hline
        $1^{+}(3^3P_1)$ &  3281.27 &  3365&  3085& 3508.63 & 3519 & 3244\\
        \hline
        $2^{+}(3^3P_2)$ &  3337.81 & 3407 &  3142& 3563.20 & 3580  & 3283\\
    \hline
   \end{tabular}
   \end{table}
    Our calculated masses are compared with available theoretical information and found that our estimated masses lies below prediction of relativistic quark model in Ref [39] with in difference of 25-50 MeV. On comparing with Ref [40], our results lies above values computed in Ref [40].
    So our computed masses for n = 3 charm mesons without and with strangness are overall in good agreement with other theoretical estimates.
    
 \subsection{Decay Widths}
    By using predicted masses, we compute decay width for n = 3 charm mesons from excited state to ground state with emission of pseudoscalar particles ($\pi,\eta,K$) only in terms of coupling constants. Input values used for calculating decay width are $M_{\pi^{0}}$ = 134.97 MeV, $M_{\pi^{+}}$ = 139.57 MeV, $ M_{K^{+}}$= 493.67 MeV, $M_{\eta^{0}}$= 547.85 MeV, $ M_{K^{0}}$= 497.61 MeV, $M_{D^{0}}$ = 1864.83 MeV, $M_{D^{\mp}}$ = 1869.65 MeV, $M_{D^{\pm}_{S}}$ = 1968.34 MeV  $ M_{D^{*0}}$ = 2006.85 MeV, $ M_{D^{*\pm}}$= 2010.26 MeV, $M_{D^{*\pm}_{S}}$= 2112.20 MeV, $M_{D^{*0}_{0}}$= 2318  MeV, $M_{D^{*\pm}_{S0}}$ = 2317.70 MeV, $M_{D^{'0}_{1}}$ = 2420.80 MeV, $M_{D^{'\pm}_{S1}}$ = 2459.50 MeV and calculated masses for n = 3 S-wave, P-wave charm mesons mentioned in Table III.\\
 The computed strong decay widths in terms of coupling constants  $\tilde{\tilde{g}}_{HH}$, $\tilde{\tilde{g}}_{SH}$, $\tilde{\tilde{g}}_{TH}$ for radially excited charm mesons are presented in Table IV and Table V. Without enough experimental data, it is not possible to determine values of coupling constants from heavy quark symmetry solely but upper bound to these coupling are mentioned in Table IV and Table V. In our study, we are taking limited modes of decays and that also only to ground state. We believe that a particular state like D(3030) give 7783.54$\tilde{\tilde{g}}_{HH}^2$ total decay width, when compared with total decay widths mentioned by other theoretical paper, we provided an upper bound on $\tilde{\tilde{g}}_{HH}^2$ value. Now if we take additional modes, then value of $\tilde{\tilde{g}}_{HH}^2$ will be lesser than 0.12 ($\tilde{\tilde{g}}_{HH}^2< 0.12$). So, these upper bounds may give important information to other associated charm states.
 Large fractions of decay width of any excited state is dominated by modes that includes ground state.  Our work also provides lower limit to total decay width which give important clues to forthcoming experimental studies.
     \begin{table*}{\normalsize
\renewcommand{\arraystretch}{1.0}
\tabcolsep 0.2cm \caption{\label{tab:expt} Decay Width of obtained masses for n=3 charm mesons.}
 \centering
   \begin{tabular}{|c|c|c|c|c|}
    \hline
    States & $J^{P}$ & Decay Modes& Decay Widths (MeV) & Upper bound of coupling constant \\
    \hline
    $D(3030.09)$ & $0^{-}$ & $D^{*\pm}\pi^-$ & 3703.68$\tilde{\tilde{g}}_{HH}^2$&\\
    & & $D^{*0}\pi^{0}$ & 1867.45$\tilde{\tilde{g}}_{HH}^2$ &\\
    & & $D^{*0}\eta^{0}$&378.62$\tilde{\tilde{g}}_{HH}^2$&\\
    & & $D^{*+}_{s}K^{-}$ & 1833.79$\tilde{\tilde{g}}_{HH}^2$&\\
    & &Total & 7783.54$\tilde{\tilde{g}}_{HH}^2$ & \\
    & & & $\tilde{\tilde{g}}_{HH}$ & 0.12 \\
     \hline
     $D(3064.45)$ & $1^{-}$ & $D^{+}\pi^-$ &  1668.5$\tilde{\tilde{g}}_{HH}^2$& \\
      & &  $D^0\pi^0$ & 841.07$\tilde{\tilde{g}}_{HH}^2$& \\
      & & $D^{*+}\pi^-$ & 2665.54$\tilde{\tilde{g}}_{HH}^2$& \\
       & & $D^{*0}\pi^0$ & 1343.38$\tilde{\tilde{g}}_{HH}^2$& \\
       & & $D^{0}\eta^0$ & 208.05$\tilde{\tilde{g}}_{HH}^2$ &\\
      & & $D^{*0}\eta^0$ & 282.52$\tilde{\tilde{g}}_{HH}^2$ &\\
         & & $D_{s}^{+}K^-$ & 1031.87$\tilde{\tilde{g}}_{HH}^2$& \\
         & & $D_{s}K^0$ & 1233.43$\tilde{\tilde{g}}_{HH}^2$ &\\
         & & $D_{s}^{+}K^-$ & 1031.87$\tilde{\tilde{g}}_{HH}^2$&\\
         & & $D_{s}^{*+}K^-$ & 1389.41$\tilde{\tilde{g}}_{HH}^2$& \\
          & &Total & 9430.27$\tilde{\tilde{g}}_{HH}^2$ &  \\
          & & & $\tilde{\tilde{g}}_{HH}$ & 0.11 \\
    \hline
    $D(3243.17)$ &  $0^+$ & $D^{+}\pi^-$ & 6893.15$\tilde{\tilde{g}}_{SH}^2$& \\
     &  & $D^{0}\pi^0$ & 3464.3$\tilde{\tilde{g}}_{SH}^2$ &\\
     &  & $D^{0}\eta^0$ & 1145.25$\tilde{\tilde{g}}_{SH}^2$& \\
     & & $D^{+}_{s}K^{-}$ &6061.58$\tilde{\tilde{g}}_{SH}^2$& \\
     & &Total & 17564.28$\tilde{\tilde{g}}_{SH}^2$&\\
     & & & $\tilde{\tilde{g}}_{SH}$ & 0.09 \\
     \hline
     $D(3356.13)$ & $1^{+}$ & $D^{*0}\pi^0$ &  3528.38$\tilde{\tilde{g}}_{SH}^2$& \\
       &  & $D^{*+}\pi^-$ &  7028.54$\tilde{\tilde{g}}_{SH}^2$ &\\
      & & $D^{*0}\eta^0$ & 1160.11$\tilde{\tilde{g}}_{SH}^2$ &\\
       & & $D^{*+}_{s}K^-$ & 6053.09$\tilde{\tilde{g}}_{SH}^2$& \\
       & &Total & 17770.12$\tilde{\tilde{g}}_{SH}^2$&\\
       & & & $\tilde{\tilde{g}}_{SH}$ & 0.08 \\
       \hline
       $D(3281.27)$ & $1^{+}$ & $D^{*0}\pi^0$ &  4258.74$\tilde{\tilde{g}}_{TH}^2$& \\
       &  & $D^{*+}\pi^-$ &  8426.74$\tilde{\tilde{g}}_{TH}^2$ &\\
      & & $D^{*0}\eta^0$ & 854.16$\tilde{\tilde{g}}_{TH}^2$ &\\
       & & $D^{*+}_{s}K^-$ & 3968.1$\tilde{\tilde{g}}_{TH}^2$& \\
       & &Total & 17507.75$\tilde{\tilde{g}}_{TH}^2$&\\
       & & & $\tilde{\tilde{g}}_{TH}$ & 0.10 \\
       \hline
       $D(3337.81)$ & $2^{+}$ & $D^{+}\pi^-$ &  2705.93$\tilde{\tilde{g}}_{TH}^2$& \\
      & &  $D^0\pi^0$ & 1367.81$\tilde{\tilde{g}}_{TH}^2$ &\\
      & & $D^{*+}\pi^-$ & 5977.98$\tilde{\tilde{g}}_{TH}^2$& \\
       & & $D^{*0}\pi^0$ & 3029.18$\tilde{\tilde{g}}_{TH}^2$& \\
       & & $D^{0}\eta^0$ & 312.75$\tilde{\tilde{g}}_{TH}^2$ &\\
      & & $D^{*0}\eta^0$ & 635.44$\tilde{\tilde{g}}_{TH}^2$ &\\
         & & $D_{s}^{+}K^-$ & 1564.54$\tilde{\tilde{g}}_{TH}^2$& \\
        & & $D_{s}^{*+}K^-$ & 3010.47$\tilde{\tilde{g}}_{TH}^2$ &\\
          & &Total & 18604.1$\tilde{\tilde{g}}_{TH}^2$&\\
          & & & $\tilde{\tilde{g}}_{TH}$ & 0.079 \\
          \hline
    \end{tabular}
    }
    \end{table*}
\begin{table*}{\normalsize
\renewcommand{\arraystretch}{1.0}
\tabcolsep 0.2cm \caption{\label{tab:expt1} Decay Width of obtained masses for n = 3 strange charm mesons.}
 \centering
   \begin{tabular}{|c|c|c|c|c|}
    \hline
    States & $J^{P}$ &  Ground state Decay Modes& Decay Widths(MeV) &  Upper bound of coupling constant  \\
    \hline
    $D_{s}^{+}(3186.5)$ & $0^{-}$ & $D^*K^+$ & 3869.08$\tilde{\tilde{g}}_{HH}^2$&\\
    & & $D^{*+}K^{0}$ & 3848.36$\tilde{\tilde{g}}_{HH}^2$ &\\
    & & $D^{*+}_{s}\eta^{0}$& 1858.32$\tilde{\tilde{g}}_{HH}^2$&\\
    & &Total & 9575.76$\tilde{\tilde{g}}_{HH}^2$ & \\
    & & & $\tilde{\tilde{g}}_{HH}$ & 0.09 \\
     \hline
     $D_{s}^{+}(3209.74)$ & $1^{-}$ & $D^{+}K^0$ &  1720.77$\tilde{\tilde{g}}_{HH}^2$& \\
      & &  $D^0K^+$ & 1739.82$\tilde{\tilde{g}}_{HH}^2$& \\
      & & $D^{*+}K^0$ & 2712.35$\tilde{\tilde{g}}_{HH}^2$& \\
       & & $D^{*0}K^{+}$ & 2744.24$\tilde{\tilde{g}}_{HH}^2$& \\
       & & $D_{s}^{+}\eta^0$ & 915.91$\tilde{\tilde{g}}_{HH}^2$ &\\
      & & $D_{s}^{*+}\eta^0$ & 1329.13$\tilde{\tilde{g}}_{HH}^2$ &\\
         & & $D_{s}^{*+}\pi^0$ & 948.81$\tilde{\tilde{g}}_{HH}^2$& \\
          & &Total & 82111.03$\tilde{\tilde{g}}_{HH}^2$ &  \\
          & & & $\tilde{\tilde{g}}_{HH}$ & 0.29 \\
          \hline
     $D_{s}^{+}(3496.46)$ & $0^{+}$ & $D_{S}^{+}\pi^0$ &  4517.26$\tilde{\tilde{g}}_{SH}^2$& \\
      & &  $D_{s}^{+}\eta$ & 6002.88$\tilde{\tilde{g}}_{SH}^2$& \\
      & & $D^{0}K^+$ & 9826.27$\tilde{\tilde{g}}_{SH}^2$ &\\
       & & $D^{+}K^0$ & 9863.43$\tilde{\tilde{g}}_{SH}^2$ &\\
      & &Total & 30209.84$\tilde{\tilde{g}}_{SH}^2$ &  \\
      & & & $\tilde{\tilde{g}}_{SH}$ & 0.06 \\
     \hline
    $D_{s}^{+}(3567.18)$ & $1^{+}$ & $D^*K^+$ & 8194.67$\tilde{\tilde{g}}_{SH}^2$&\\
    & & $D^{*+}K^{0}$ & 8179.21$\tilde{\tilde{g}}_{SH}^2$ &\\
    & & $D^{*+}_{s}\eta^{0}$& 5723.21$\tilde{\tilde{g}}_{HH}^2$&\\
    & & $D_{s}^{+*}\pi^0$ &  4333.65$\tilde{\tilde{g}}_{SH}^2$& \\
    & &Total & 26430.74$\tilde{\tilde{g}}_{SH}^2$ & \\
    & & & $\tilde{\tilde{g}}_{SH}$ & 0.07 \\
    \hline
    $D_{s}^{+}(3508.63)$ & $1^{+}$ & $D^*K^+$ & 9302.31$\tilde{\tilde{g}}_{TH}^2$&\\
    & & $D^{*+}K^{0}$ & 9246.22$\tilde{\tilde{g}}_{TH}^2$ &\\
    & & $D^{*+}_{s}\eta^{0}$& 5675.62$\tilde{\tilde{g}}_{TH}^2$&\\
    & & $D_{s}^{+*}\pi^0$ &  7613.48$\tilde{\tilde{g}}_{TH}^2$& \\
    & &Total & 31837.63$\tilde{\tilde{g}}_{TH}^2$ & \\
    & & & $\tilde{\tilde{g}}_{TH}$ & 0.06 \\
    \hline
          $D_{s}^{+}(3563.20)$ & $2^{+}$ & $D^*K^+$ & 16804.3$\tilde{\tilde{g}}_{TH}^2$&\\
    & & $D^{*+}K^{0}$ & 16885.1$\tilde{\tilde{g}}_{TH}^2$ &\\
    & & $D^{*+}_{s}\eta^{0}$& 5675.62$\tilde{\tilde{g}}_{TH}^2$&\\
    & & $D_{s}^{*+}\pi^0$ &  1150.23$\tilde{\tilde{g}}_{TH}^2$& \\
     &  & $D_{s}^{+}\eta$ & 11505.4$\tilde{\tilde{g}}_{SH}^2$& \\
      & & $D^{+}K^0$ & 22261.8$\tilde{\tilde{g}}_{SH}^2$ &\\
       & & $D^{0}K^+$ & 22204.4$\tilde{\tilde{g}}_{SH}^2$ &\\
     & & $D_{s}^{+}\pi^0$ & 11872.7$\tilde{\tilde{g}}_{TH}^2$& \\
    & &Total & 108359.38$\tilde{\tilde{g}}_{TH}^2$ & \\
    & & & $\tilde{\tilde{g}}_{TH}$ & 0.035 \\
          \hline
    \end{tabular}
    }
    \end{table*}
    \subsection{Regge Trajectory}
    Regge trajectories are a powerful tool to study spectroscopy of hadron spectra. The graph between total angular momentum(J) and radial quantum number($n_r$) of hadrons against square of their masses($M^2$) gives information about quantum number of particular state and also helps to identify recently observed states. We use following definitions:
    \begin{itemize}
        \item[(a).]{The (J,$M^2$ ) Regge trajectories:
    \begin{equation}
        J = \alpha M^2 + \alpha_{0}
      \end{equation}}
      \item[(b).]{The ($n_r$, $M^2$ ) Regge trajectories:
    \begin{equation}
       n_r  = \beta M^2 + \beta_{0}
    \end{equation} }
    \end{itemize}
      here $\alpha$, $\beta$ are slopes and $\alpha_{0}$, $\beta_{0}$ are intercepts. We construct Regge trajectories in plane (J, $M^2$ ) with natural parity $P= (-1)^J$ and unnatural parity $P= (-1)^{J-1}$ depicted in fig. $1-4$. Regge trajectories in plane ($n_r$, $M^2$) are constructed in fig. $5-6$ using spin averaged masses for S and P waves where spin averaged masses for S-wave is given by $S = (3m^{Q}_{H^*}+m^{Q}_{H})/8$  and for P-wave is $P = (5m^{Q}_{S^*}+3m^{Q}_{S})/8$. In fig $1-6$, masses for n = 1 are taken from PDG; for n = 2 masses are taken from Ref. [39] and for n = 3, we are taking our calculated masses. Our calculated data fit on Regge lines with good accuracy. By fixing slopes and intercepts of these Regge trajectories (Table VI), we calculate higher masses listed in Table VII and VIII. Using Regge trajectories, we also assign quantum number to state $D_{2}^{*}(3000)$. The state $D_{2}^{*}(3000)$ is reported by LHCb collaboration in 2016 by studying $B^{-}\rightarrow D^+ \pi^- \pi^-$ decay . This state is analyzed by different theoretical models and suggested different assignment. Using relativistic quark model, Zhi-Gang Wang studied $D_{2}^{*}(3000)$ state and assign it to be $1F^{+}_\frac{5}{2}$ state [28]. They also analyzed this state by help of $^{3}P_{0}$ model and suggested it to be 3$^{3}P_{2}$. There is an ambiguity in $J^P$ of the state  $D_{2}^{*}(3000)$. So, here we assigned proper $J^P$ for $D_{2}^{*}(3000)$ with Regge trajectories in plane (J, $M^2$) and suggested it to be $1F(2^{+})$ state shown in fig. 7. In fig.7, mass for $1^3P_0$ is taken from PDG [11]; mass for $1^3D_1$ is from Ref [39] and we take $D_{2}^{*}(3000)$ as  $1^3F_2$.
  \begin{figure}[htp]
    \centering
\subfloat[Regge trajectories for non strange charm meson with unnatural parity\label{fig:1a}]{ \includegraphics[width=6cm]{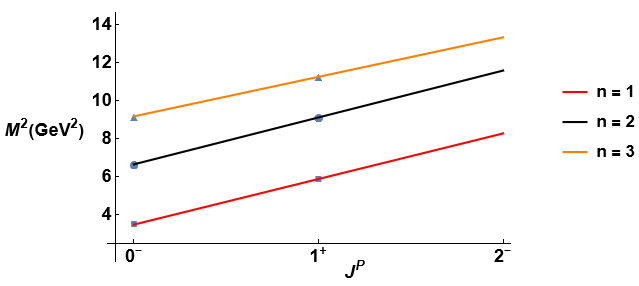}}\\
\subfloat[Regge trajectories for non strange charm meson with natural parity\label{fig:1b}]{
    \includegraphics[width=6cm]{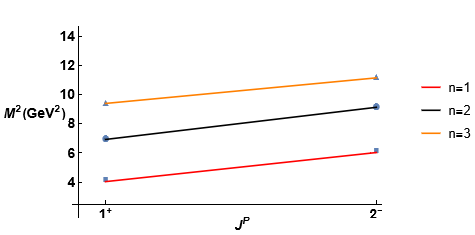}}
    
\end{figure}

\begin{figure}[htp]
    \centering
    \includegraphics[width=9cm]{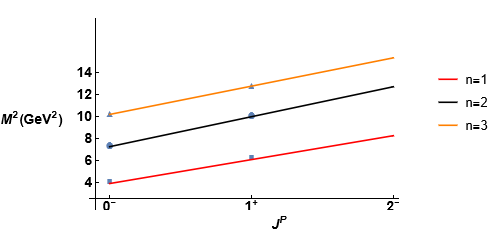}
    \caption{Regge trajectories for strange charm mesons with unnatural parity }
    \label{fig:my_label3}
\end{figure}
\begin{figure}[t!]
    \centering
    \includegraphics[width=9cm]{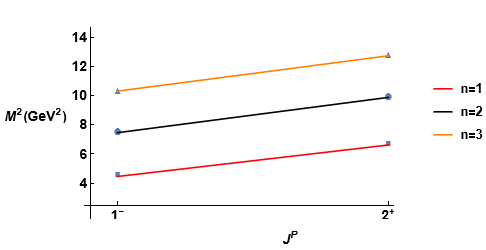}
    \caption{Regge trajectories for strange charm meson with natural parity }
    \label{fig:my_label4}
\end{figure}
\begin{figure}[htp]
    \centering
    \includegraphics[width=9cm]{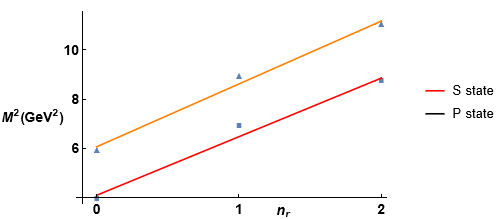}
    \caption{Regge trajectories for spin average masses for non strange charm meson in plane($M^2\rightarrow n_r$)}
    \label{fig:my_label5}
\end{figure}
\begin{figure}[htp]
    \centering
    \includegraphics[width=9cm]{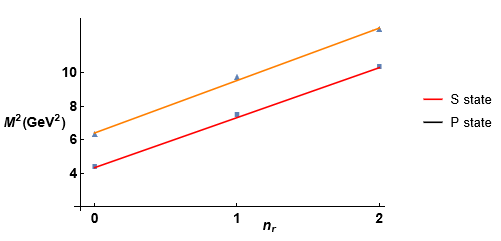}
    \caption{Regge trajectories for spin average masses for strange charm meson in plane($M^2\rightarrow n_r$)}
    \label{fig:my_label6}
\end{figure}

\begin{figure}[htp]
    \centering
    \includegraphics[width=9cm]{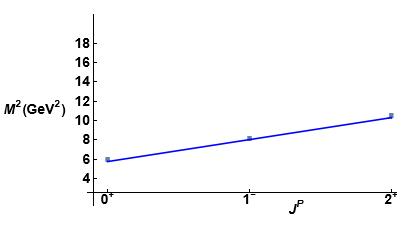}
    \caption{Regge lines in plane($M^2\rightarrow J^P$) to identify $D_{2}(3000)$}
    \label{fig:my_label7}
\end{figure}

\begin{table}[htp!]
    \centering
    \begin{tabular}{|c|c|c|c|c|}
    \hline
       Mesons  & State & Slope($\beta$) & Intercepts($\beta_{0}$)\\
         \hline
         & $0^{-}$ & 0.350235 & -1.25767\\
         &  $1^{-}$ & 0.372595 & -1.53582\\
         & $0^{+}$ & 0.41673& -2.45654\\
          D & $1^{+}$ & 0.37199 & -2.18897\\
         & $1^{+}$ & 0.404165 & -2.42289\\
         & $2^{+}$ & 0.92575& -2.39406\\
         \hline
          & $0^{-}$ & 0.317195 & -1.18762\\
         &  $1^{-}$ & 0.339684 & -1.45611\\
         & $0^{+}$ & 0.291202& -1.53521\\
          $D_{S}$ & $1^{+}$ & 0.299011 & -1.78088\\
         & $1^{+}$ & 0.332348& -2.02826\\
         & $2^{+}$ & 0.325032& -2.07554\\
         \hline
    \end{tabular}
    \caption{Regge Slopes and Regge Intercepts }
    \label{tab:my_label60}
\end{table}
\begin{table}[htp!]
\centering
    \begin{tabular}{|c|c|c|c|}
    \hline
     State  & $J^{P}$ & Masses (MeV) & Ref[25] \\
     \hline
        $4^{1}S_{0}$ & $0^{-}$ & 3484.65& 3468\\
        \hline
        $4^{3}S_{1}$ & $1^{-}$ & 3489.07& 3497\\
        \hline
        $4^{3}P_{0}$ & $0^{+}$ & 3618.52& 3697\\
        \hline
         $4^{1}P_{1}$ & $1^{+}$& 3733.82& 3709\\
         \hline
        $4^{3}P_{1}$ & $1^{+}$& 3662.99& 3681\\ 
        \hline
        $4^{3}P_{2}$ & $1^{+}$& 3716.67& 3701\\
        \hline
    \end{tabular}
    \caption{Higher Non Strange Masses Lying in Regge lines in plane $(n_{r},M^{2})$}
    \label{tab:my_label30}
\end{table}
\vspace{-3.0cm}
\begin{table}[htp]
\centering
    \begin{tabular}{|c|c|c|c|}
    \hline
     State  & $J^{P}$ & Masses (MeV) & Ref[25] \\
     \hline
        $4^{1}S_{0}$ & $0^{-}$ & 3633.46& 3547\\
        \hline
        $4^{3}S_{1}$ & $1^{-}$ & 3621.93& 3575\\
        \hline
        $4^{3}P_{0}$ & $0^{+}$ & 3946.4& 3764\\
        \hline
         $4^{1}P_{1}$ & $1^{+}$& 3998.62& 3778\\
         \hline
        $4^{3}P_{1}$ & $1^{+}$& 3889.67& 3764\\ 
        \hline
        $4^{3}P_{2}$ & $1^{+}$& 3951.65& 3783\\
        \hline
    \end{tabular}
    \caption{Higher Strange Masses Lying in Regge lines in $(n_{r},M^{2})$}
    \label{tab:my_label31}
\end{table}
\vspace{5cm}
\section{Conclusion}
Heavy quark symmetry is one of important tool to describe spectroscopy of hadrons containing single heavy quark. Using available experimental as well as theoretical data on bottom mesons and applying heavy quark symmetry, we predicted masses for n=3 charm mesons spectra. With computed masses for n=3 charm mesons, we analyzed decay widths from excited state to ground state with emission of pseudoscalar mesons and expressed decay widths in terms of coupling constants. These coupling constants are computed on comparing our decay widths with  theoretical available total decay widths. The total decay widths may give upper bound on these coupling constants hence providing important clue to other associated states of charm mesons. Using our calculated charm masses for n = 3, we construct Regge trajectories in (J,$M^2$ ) and ($n_r$, $M^2$ ) planes. These Regge lines are almost linear, parallel and equidistant. Most of our predicted data nicely fit to them. We also computed masses for n = 4 charm spectra by fixing Regge slopes and intercepts in plane ($n_r$, $M^2$ ). Our calculated masses and upper bounds findings may help experimentalists for looking into higher excited states.

\section{Acknowledgment}
The authors thankfully acknowledge the financial support by the
Department of Science and Technology (SERB/F/9119/2020), New
Delhi.


\begin{thebibliography}{}
\bibitem{1} R. Aaij et al. (LHCb Collaboration), Phys. Rev. D 94,072001 (2016).
\bibitem{2} R. Aaij et al. (LHCb Collaboration), Phys. Rev. D 92, 32002 (2015).
\bibitem{3} R. Aaij et al. (LHCb Collaboration), JHEP 09, 145 (2013).
\bibitem{4} P. del Amo Sanchez et al. (BABAR Collaboration) Phys.Rev. D 82, 111101(R) (2010).
\bibitem{5} R. Aaij et al. (LHCb Collaboration), Phys. Rev. Lett. 126, 122002 (2021).
\bibitem{6} R. Aaij et al. (LHCb Collaboration), Eur.Phys.J.C 81, 601 (2021).
\bibitem{7} T. Aaltonen et al. (CDF Collaboration) Phys. Rev. D 90, 012013(2014)
\bibitem{8} R. Aaij et al. [LHCb Collaboration], JHEP 1504, 024 (2015)
\bibitem{9} R. Aaij et al. (LHCb), Phys. Rev. Lett. 118, 182001 (2017).
\bibitem{10} R. Aaij et al. (LHCb), Phys. Rev. Lett. 119, 112001 (2017).
\bibitem{11} P.A. Zyla, et al [Particle Data Group], Prog. Theor. Exp. Phys. 2020, 083C01 (2020)
\bibitem{12} S.Godfrey, et al , Phys. Rev. D 93, 034035 (2016)
\bibitem{13} Pallavi Gupta and A. Upadhyay, Phys. Rev. D 97, 014015(2018)
\bibitem{14} S. Narison, Phys. Lett. B 605, 319 (2005)
\bibitem{15} S. Godfrey and N. Isgur, Phys. Rev. D 32, 189(1985).
\bibitem{16} M. Di Pierro and E. Eichten, Phys. Rev. D 64, 114004(2001).
\bibitem{17} Z. G. Wang, Phys. Rev. D 83, 014009 (2011)
\bibitem{18} P. Colangelo, F. De Fazio and S. Nicotri, Phys. Lett. B 642,48 (2006)
\bibitem{19} P. Colangelo, F. De Fazio, F.Giannuzzi and S. Nicotri, Phys. Rev. D 86 ,054024 (2012) .
\bibitem{20} A. M. Badalian and B. L. G. Bakker, Phys. Rev. D 84, 034006(2011).
\bibitem{21} Q. F. Lu and D. M. Li, Phys. Rev.D 90, 054024(2014).
\bibitem{22} Q. T. Song, D. Y. Chen, X. Liu and T. Matsuki, Phys. Rev. D 92, 074011 (2015).
\bibitem{23} B.Chen, X. Liu and A. Zhang, Phys. Rev. D 92, 034005 (2015).
\bibitem{24} Z. G. Wang, Phys. Rev. D 88, 114003(2013).
\bibitem{25} S. Godfrey and K. Moats, Phys. Rev. D 93, 034035 (2016).
\bibitem{26} P. del Amo Sanchez et al, Phys. Rev. D 82, 111101 (2010).
\bibitem{27} R. Aaij et al, JHEP 1309, 145 (2013).
\bibitem{28} Guo Liang Yu, Z.G.Wang, Phys. Rev. D 94, 074024 (2016)
\bibitem{29} S. Godfrey and K. Moats, Phys. Rev. D 93, 034035 (2016).
\bibitem{30} Y. Sun, X. Liu, and T. Matsuki, Phys. Rev. D 88, 094020 (2013).
\bibitem{31} M. Neubert,Phys. Rept.245,259 (1994).
\bibitem{32} A.Falk and T. Mehen, Phys. Rev. D 53 231 (1996).
\bibitem{33}  Zhi-Gang Wang, Eur.Phys.J.Plus 129, 186 (2014).
\bibitem{34} M. B.Wise, Phys. Rev. D 45, 2188 (1992). 
\bibitem{35} G. Burdman and J. F. Donoghue, Phys. Lett. B 280, 287 (1992).
\bibitem{36} P. L. Cho, Phys.Lett. B 285, 145 (1992).
\bibitem{37} Phys. Rev. D 46, 1148 (1992) [Erratum-ibid. D 55, 5851 (1997)].
\bibitem{38} R. Casalbuoni, A. Deandrea,N. Di Bartolomeo, R. Gatto, F. Feruglio and G. Nardulli, Phys. Lett. B 299, 139 (1993).
\bibitem{39} D. Ebert, R. N. Faustov and V. O. Galkin, Eur. Phys. J.C66, 197  (2010).
\bibitem{40} T. A. Lahde, C. Nyfalt, D. O. Riska, 	Nucl.Phys. A674, 141-167 (2000).
\end{thebibliography}
\end{document}